\documentclass[sigconf]{acmart}

\AtBeginDocument{%
  \providecommand\BibTeX{{%
    \normalfont B\kern-0.5em{\scshape i\kern-0.25em b}\kern-0.8em\TeX}}}
    
\usepackage{multirow}
\usepackage{mathtools}
\usepackage{subcaption}
\usepackage{balance}
\usepackage{color}
\usepackage{wrapfig,acmart-taps}
\usepackage{float}

\copyrightyear{2023}
\acmYear{2023}
\setcopyright{rightsretained}
\acmConference[FAccT '23]{2023 ACM Conference on Fairness, Accountability,
and Transparency}{June 12--15, 2023}{Chicago, IL, USA}
\acmBooktitle{2023 ACM Conference on Fairness, Accountability, and
Transparency (FAccT '23), June 12--15, 2023, Chicago, IL, USA}
\acmDOI{10.1145/3593013.3594115}
\acmISBN{979-8-4007-0192-4/23/06}

\begin{document}
\title{Discrimination through Image Selection by Job Advertisers on Facebook}

\author{Varun Nagaraj Rao}
\affiliation{%
  \institution{Princeton University}
  \city{Princeton}
  \state{NJ}
  \country{USA}
  }
\email{varunrao@princeton.edu}

\author{Aleksandra Korolova}
\affiliation{%
  \institution{Princeton University}
  \city{Princeton}
  \state{NJ}
  \country{USA}
  }
\email{korolova@princeton.edu}

\begin{abstract}
Targeted advertising platforms are widely used by job advertisers to reach potential employees; thus issues of discrimination due to targeting that have surfaced have received widespread attention. 
Advertisers could misuse targeting tools to exclude people based on gender, race, location and other protected attributes from seeing their job ads. 
In response to legal actions, Facebook disabled the ability for explicit targeting based on many attributes for some ad categories, including employment.
Although this is a step in the right direction, prior work has shown that discrimination can take place not just due to the explicit targeting tools of the platforms, but also due to the impact of the biased ad delivery algorithm. Thus, one must look at the potential for discrimination more broadly, and not merely through the lens of the explicit targeting tools.

In this work, we propose and investigate the prevalence of a new means for discrimination in job advertising, that combines both targeting and delivery -- through the disproportionate representation or exclusion of people of certain demographics in job ad images. 
We use the Facebook Ad Library to demonstrate the prevalence of this practice through: (1) evidence of advertisers running many campaigns using ad images of people of only one perceived gender, (2) systematic analysis for gender representation in all current ad campaigns for truck drivers and nurses, (3) longitudinal analysis of ad campaign image use by gender and race for select advertisers.
After establishing that the discrimination resulting from a selective choice of people in job ad images, combined with algorithmic amplification of skews by the ad delivery algorithm, is of immediate concern, we discuss approaches and challenges for addressing it. 
\end{abstract}


\maketitle

\section{Introduction}
Targeted advertising platforms are widely used by job advertisers to reach potential employees. For example, a study commissioned by Facebook in 2018 found that one in four people in the U.S. searched for, or found a job using Facebook's platform \cite{fbjobs2018importance}. At the same time, issues of discrimination due to targeting have surfaced and received widespread attention in recent years. Advertisers could misuse the targeting tools, including algorithmic audience creation mechanisms, to exclude people based on gender, race, location and other legally protected attributes from seeing their job ads \cite{fb3, hud2019facebook, propub2018men, propub2017age, sapiezynski2022algorithms, faizullabhoy2018facebook}. In response to lawsuits by the American Civil Liberties Union to the U.S. Equal Employment Opportunity Commission \cite{aclu2018lawsuit}, and Facebook's commissioned Civil Rights Audit~\cite{fb29}, Facebook in 2019 disabled the explicit targeting based on gender, age and zip code for some ad categories, including employment \cite{fb1}. 

Although this is a step in the right direction, researchers \cite{ali2019discrimination, imana2021auditing} and the U.S. Department of Justice \cite{doj2022fb} demonstrate that discrimination can take place not just due to the explicit targeting tools made available by the platforms, but also due to the impact of a bias in the ad delivery algorithm. Thus, one must look at the potential for discrimination broadly, beyond the lens of the explicit targeting.  

An important and to-date, under-examined question is the potential for discrimination through the disproportionate representation of certain people in job ad images. Selective image use can lead to discrimination as follows:
First, images have a unique persuasive power and social science literature~\cite{walker2012diversity, white2019role, thaler2001diversify, avery2006target} demonstrates images present on recruitment websites affect the application intention of job seekers and can be used to manipulate the gender and racial composition of those who apply. 
Second, complementary evidence from case law~\cite{johnson1991fair} shows that portraying only the dominant demographic in images discourages minorities from seeking out those opportunities. 
And third, a skew in image selection can be algorithmically amplified by ad delivery algorithms, solely based on the demographic characteristics of the people depicted~\cite{kaplan2022measurement}. 

To investigate the prevalence of such discrimination, that combines both targeting and delivery in online advertising, we analyze job advertisers' selective use of people in job ad images on Facebook. 
We hypothesize that a job advertiser attempting to exclude or discourage individuals of a certain gender from applying to their jobs (or equivalently, to make the job more appealing to applicants of one gender, and less appealing to applicants of another gender), thereby circumventing the targeting restrictions of the platform, can do so through the selective use of people in the images chosen for their job ad campaigns. For example, a trucking company interested in hiring only men as drivers, could create job ads using images containing only men in them.  
Thus, we point out that in modern ad systems, a job advertiser's selective use of images depicting people of only one demographic is almost analogous to explicitly targeting that demographic.  

We collect and analyze data\footnote{All data used in our analysis can be accessed at \url{https://github.com/varunnrao/job_ad_images_facct23}} from the Facebook Ad Library to demonstrate that job advertisers are already leveraging the selective use of people in job ad images for potential discrimination through:
\begin{itemize}
\item Evidence of advertisers running many campaigns using ad images of people of only one perceived gender.
\item Systematic analysis for perceived gender representation in all current ad campaigns for truck drivers and nurses.
\item Analysis of perceived gender and race representation for select advertisers.
\end{itemize}

After establishing that the discrimination, resulting from the selective choice of people in job ad images, combined with algorithmic amplification of skews thanks to ad delivery optimization, is of immediate concern, we discuss approaches for addressing it. 
Specifically, we describe the data and functionality which should be (but currently is not) provided by the ad platforms to enable public-interest researchers to detect potentially discriminatory image selection targeting. 
Furthermore, we underscore the necessity of transparency not only with regards to the advertiser choices, but also with regards to the ad delivery algorithm as applied to employment advertising.
We then draw parallels to discrimination through selective use of language, suggesting that selective use of people in images should have similar legal protections and platform guidance.
Finally, drawing on literature from jury representation and special education, we discuss approaches for measuring advertiser intent in image selection, and the challenges to their adoption.

In summary, our main contributions are:\\
(1) define a new means of discriminative targeting by advertisers through selective use of people in the ad images, and demonstrate its prevalence for job ads on Facebook (Section \ref{sec:rq1}). \\
(2) provide desiderata for platform's actions to address our findings and discuss the challenges of establishing normative metrics on the basis of advertiser choices alone (Section~\ref{sec:solutions}).
\section{Background and Related Work}
\label{sec:related_work}
Before we introduce our methods and analyses, we provide background and related work on discrimination in targeted advertising systems and the corresponding legal, policy and platform responses to previously discovered issues. 
Subsequently, we highlight the unique role played by images in employment advertising and motivate the need to study job advertiser image selection.

\subsection{Ad Targeting, Delivery and Discrimination}
\subsubsection{The ad system: targeting and delivery}

Facebook's ad system consists of two phases - ad creation and ad delivery~\cite{fb16, fb17}. 
During \textit{ad creation} the advertiser chooses 
(1) their business objective, i.e. an outcome they'd like to achieve, such as increasing the number of visitors to their website, 
(2) contents of the ad including the images, text, and destination page link, 
(3) the targeting parameters, specifying the kinds of users they'd like the ad to be shown to, and 
(4) the budget and duration of their ad campaign.
Advanced targeting capabilities, using demographic, geographic, interest and/or behavioral characteristics of users have long been touted by Facebook as a useful tool for advertisers.

\textit{Ad delivery} is the process through which a subset of the users targeted by the advertiser are chosen to see the ad.
For each user, the platform runs an auction among all advertisers targeting that user, to determine which ad to show.
The high-level information publicly known about the auction is that it does not necessarily select the ad with the highest bid; rather, for each candidate ad, Facebook computes its \textit{Total Value}, defined as:\\
$
\textit{Total Value = } \textit{Advertiser Bid} \times \textit{Estimated Action Rate} + \textit{Ad Quality};
$
and then shows the user the ad with the highest total value~\cite{fb2023vrspaper, fb16, fb17}.
All components of this equation are computed using machine learning. 
The \textit{Estimated Action Rate} is Facebook's predicted likelihood of the user taking an advertiser's desired action, 
the \textit{Ad Quality} is a machine learning model that combines feedback from users and predictions based on image and text of the ad creative, and even the \textit{Advertiser Bid} is an algorithmic model, where Facebook chooses what to bid on behalf of the advertiser based on their budget and expressed preference in bid strategy~\cite{fb24B}. 

\subsubsection{Discrimination in ad targeting} 
\label{sec:target_discrimination}
Prior work has found significant evidence of Facebook's extensive targeting tools being used in order to exclude individuals on the basis of gender, age, or race in housing and employment advertising (which are areas governed by U.S. anti-discrimination laws). For example, Facebook tools offered the ability to exclude Black and Hispanic users living in a specific area from the targeting of housing ads \cite{fb3, hud2019facebook}. Furthermore,
\cite{propub2018men} found ads from ten traditionally male-dominant industries targeted just at men, including a software company, a moving company and a police department. Moreover, \cite{propub2017age} found that advertisers excluded older workers from seeing job ads. For instance, financial analyst job ads were only targeted at users aged 25--36 years. 

\subsubsection{Discrimination in ad delivery}
\label{sec:delivery_discrimination}
Discrimination can occur not only due to advertisers' choices during targeting, but also as a result of the ad delivery process, as the machine learning driven components of the total value equation can be biased. 

Specifically, \citet{ali2019discrimination} showed that in the case of Facebook, the ad creative image influences and skews the ad delivery along gender and racial lines, in ways that cannot be explained by market effects or users' interactions with the ads. 
Follow-up work by \citet{imana2021auditing} demonstrated that the skew in ad delivery by gender in the case of employment ads cannot be justified by differences in qualifications among demographics in the target audience, and is thus fully attributable to algorithmic decisions made by Facebook in its own business interests. 
More recently, concurrent and complementary work by \citet{kaplan2022measurement} showed that ad delivery can be dramatically skewed solely as a result of demographic characteristics of the people depicted in the images. 
These works provide evidence of discrimination in ad delivery, a hypothesis that was put forward by~\citet{sweeney2013discrimination} and further investigated by~\cite{datta2015automated, datta2018discrimination, lambrecht2019algorithmic, kingsley2020auditing, celis2019toward, dwork2018fairness}. 

\subsection{Legal, Policy and Platform Response}
\subsubsection{Removal of exclusionary targeting features}
In response to the investigative reporting~\cite{fb3, propub2018men,propub2017age}, lawsuits~\cite{hud2019facebook, aclu2018lawsuit},
and a Civil Rights Audit~\cite{fb29}, Facebook disabled the advertiser ability to explicitly target employment ads by age or gender in  2019~\cite{fb1}. 
The ability to explicitly target by race and ethnicity had been removed earlier in 2018 \cite{fb34}; however, proxies for targeting by racial categories remained available even in 2022 \cite{keegan2021race, fb35}.
Facebook also introduced a requirement for advertisers to self-identify when running ads on housing, employment, or credit (HEC) issues, and to acknowledge adherence to Facebook's Discriminatory Practices policy~\cite{fb14, fb2023discriminatoryprac}.
Google~\cite{google2020disable} followed suit by restricting targeting criteria for HEC ads and requiring advertisers to acknowledge adherence to a personalized advertising policy. There is no advertiser-facing feature on Google to self-identify the ad category; instead, ads may be automatically labeled as belonging to an HEC category after creation and review.  LinkedIn retained the ability for advertisers to target based on age or gender, but required advertisers to self-certify that for their HEC or education ads, they will not use the platform to discriminate based on age or gender~\cite{linkedin2021disable}.

\subsubsection{Progress on reducing discrimination in ad delivery}
Addressing discrimination in ad delivery remains more complex.
In a June 2022 settlement with the Department of Justice, Facebook committed to ``develop a new system to address racial and other disparities caused by its use of personalization algorithms in its ad delivery system'', called the Variance Reduction System (VRS)~\cite{doj2022fb}.
Specifically, the VRS system, aims to ensure that the age, gender and estimated race distribution of the audience that is shown a housing ad closely resembles the distribution of age, gender and estimated race of the audience targeted by the advertiser~\cite{fb2023vrspaper}.
During an ad's delivery, Facebook periodically compares the ratio of impressions delivered to a particular age / gender / racial subgroup, with that subgroup's fraction in the advertiser's target audience. 
When the ratios diverge, Facebook adjusts one of the machine learning algorithms used to calculate the ad's total value in the auction, in a way that will change the likelihood that this ad will win the auction and be shown to a user of a particular subgroup.
Although as of January 2023, the VRS system is implemented only for housing ads, Facebook has committed to expanding it also to employment advertising \cite{fb2023vrs}. 

A limitation of the VRS approach for eliminating discrimination is that the demographic composition of the advertiser's targeted audience is chosen as the goal for the composition of the delivery audience. Such a choice may be problematic as long as the advertisers continue to have access to tools (e.g. custom audiences, location based audience creation) to select their targeted audience in a discriminatory manner~\cite{speicher2018potential, sapiezynski2022algorithms, faizullabhoy2018facebook}.

As will become clear from subsequent discussion in Sections~\ref{sec:ss} and~\ref{sec:cl}, the VRS approach is unlikely to address the new type of discrimination through image selection that we identify in this work. Extrapolating from the social science literature and case law~\cite{walker2012diversity, white2019role, thaler2001diversify, avery2006target, johnson1991fair}, the selective use of people in ad images affects the ad recipients' likelihood of acting on the ad (e.g., their likelihood of \textit{clicking} on it), whereas the VRS is entirely focused on minimizing the variance in \textit{impressions} (i.e. showings of the ad).

\subsubsection{The Facebook Ad Library and its functionality.}
The \textit{Facebook Ad Library}\footnote{\url{https://www.facebook.com/ads/library/}} is a transparency tool created by Facebook that provides a keyword-based interface to a searchable collection of active ads on the platform. After choosing an ad category among (1) all ads, (2) housing, (3) employment, (4) credit, or (4) issues, elections or politics (referred to as political ads henceforth), one can search for ads based on specific keywords or suggested  advertiser names. 

The search results page loads ads in the category chosen in reverse chronological order as one scrolls down the page. 
For all active ads, the search results page contains details about the ad content used, including one or more of the ad text, destination page link, video, and image(s), and other meta data such as the Ad ID, and the date it started running. 
However, it does not contain data about inactive ads. 
Critically, with the exception of political ads, neither the ad targeting information (such as advertiser targeting choices or campaign budget) nor the ad delivery information (such as the number of people an ad reached, location where it was delivered, or a demographic breakdown of the ad recipients) is available.

Most prior works using the Facebook Ad Library focused on political and issue advertising \cite{edelson2020security, le2022audit, silva2020facebook, sosnovik2021understanding}. We are aware of only one effort that uses the data in the ad library for inferences about employment advertising --- a class action charge by Real Women in Trucking against Meta (formerly Facebook) alleging that Meta discriminates against women and older people when deciding which users receive employers' job ads on Facebook~\cite{truck2022lawsuit}. 
In it, the supporting evidence lists over 75 employment ads from the Ad Library that were delivered primarily to men. For example, an employer seeking truck drivers in North Carolina reached an audience that was only 5\% female.

\subsection{Role of Images for Job Seekers}
Our work is focused on the potential ability and practice of discrimination through image selection. We now describe why images play a unique role in employment advertising from the social science and policy perspectives.

\subsubsection{The social science perspective}\label{sec:ss}
Images play a unique role in advertising due to their distinctive characteristics, which are not present in other sources of media \cite{messaris1997visual, smith2008visual}. 
In his 1997 book ``Visual Persuasion: The Role of Images in Advertising'' \cite{messaris1997visual}, Paul Messaris outlines iconicity, indexicality, and syntactic indeterminacy as the fundamental traits of images that distinguish them from other modalities and uniquely affect the ad campaign. 
First, the iconicity of images gives advertisers access to strong emotional responses invoked in the target audience when they view an image. 
Second, images are indexical and can serve as documentary evidence of an advertiser's point of view. 
Third, images lack explicitness and syntax, and thus the audience's interpretation of the message conveyed through an image in an ad campaign is a creation of their own. 

Based on social science theories, researchers argue that images present on recruitment websites and in advertising affect the application intention of job seekers and can be used to manipulate the gender and racial composition of those who apply. 
Recent works \cite{walker2012diversity, white2019role} applied Spence's Signaling Theory~\cite{spence1973job}, Social Identity Theory~\cite{tajfel1974social} and Visual Perception Theory~\cite{gibson2002theory} in support of this hypothesis. 
Based on the Signaling and Perception Theories, \cite{white2019role} argued that people who apply for jobs will resemble in their demographic characteristics those of the people depicted on the job websites. 
In an apparent attempt to leverage the effects of these theories in practice, organizations have been found to use images of diverse people in job ads to recruit diverse talent \cite{thaler2001diversify, avery2006target}. 
In parallel, \cite{walker2012diversity} leveraged the Social Identity Theory to deduce that ``individuals naturally categorize themselves and others in terms of important visible characteristics (e.g., race, gender, age)'', and ``develop more favorable attitudes toward in-group members and seek environments that affirm their identity''. As a result, they demonstrate that job seekers infer diversity cues from the people present in the images, which in turn influences their job application decision. 

\subsubsection{Policy context}\label{sec:cl}
Legal scholarship under both Section 704(b) of Title VII \cite{usc1964TitleVII} and the parallel, but broader Fair Housing Act (FHA) \cite{usc1968FHA} of U.S. law suggests that use of images containing only people of the dominant demographic in jobs or housing ads may be interpreted by the casual observer to convey a preference based on a protected class, irrespective of the advertiser's intent \cite{1983capaci, datta2018discrimination}. 
Evidence of such discrimination is detailed in the FHA case law \cite{johnson1991fair}. 
According to \cite{datta2018discrimination}, ``the FHA case law connects the ordinary reader standard to the prohibition on sex-designated advertising columns (e.g., those with ``Male Help Wanted'' and ``Female Help Wanted'' headings) by explaining that advertisements that exclusively feature White models may discourage Black people from pursuing housing opportunities by conveying a racial message in much the same way that the sex-designated columns furthered illegal employment
discrimination'' (see Section \ref{sec:text} for additional discussion).

FHA case law can reasonably be extended to apply to the context of employment ads as evidenced by a high profile settlement involving Abercrombie \& Finch in 2004 \cite{illston2004gonzalez}. 
As part of that settlement, Abercrombie agreed to add more diversity to their marketing materials and reflect the makeup of the nation's population; so as not to discourage members of minorities from applying for jobs that were otherwise dominated by ``well-known collegiate, all-American - and largely White'' people~\cite{nytimes2004abercrombie}.

\subsubsection{Use of images to drive a message of bias in non-advertising contexts}
The influence of images in shaping biased perceptions or exaggerating diversity has been extensively researched across various contexts beyond online advertising. 
\citet{moriearty2009framing} argued that the over-representation of Black youth as criminals in images used by media has resulted in disproportionately stricter policing and harsher sentencing for Black youth. 
Similarly, in an effort to attract under-represented members and create a misguided sense of belonging, universities have been found to over-represent African American students in their admissions brochures \cite{pippert2013we}. 
Netflix too has been accused of using misleading visual representations to entice viewers based on race; for example, some marketing posters contained only Black people even if they appeared only briefly in a show \cite{netflix2018}. 
In fact, a 12 year longitudinal study of TV shows by \citet{baruah2022tv} found that male characters with a light skin tone occupy the majority of the screen time.
Biases have also been identified in reports on vaccination and antimicrobial resistance; with White people shown in a professional capacity whereas children of color shown as vulnerable and exposed \cite{charani2023use}. 
In the Computer Science community, the works of \cite{metaxa2021image, papakyriakopoulos2021beyond, kay2015unequal} found a gender- and race- driven representation bias in both image- and text-based Google Image Search results. 
Similarly, \citet{luccioni2023stable} found that Text-to-Image models depict care professions (e.g., dental assistant) nearly exclusively through as women, and positions of authority (e.g., director, CEO) -- exclusively through men.
\section{Job Advertisers' Selective Use of People in Job Ad Images}
\label{sec:rq1}

\begin{figure*}[ht]
    \centering
    \includegraphics[width=\textwidth]{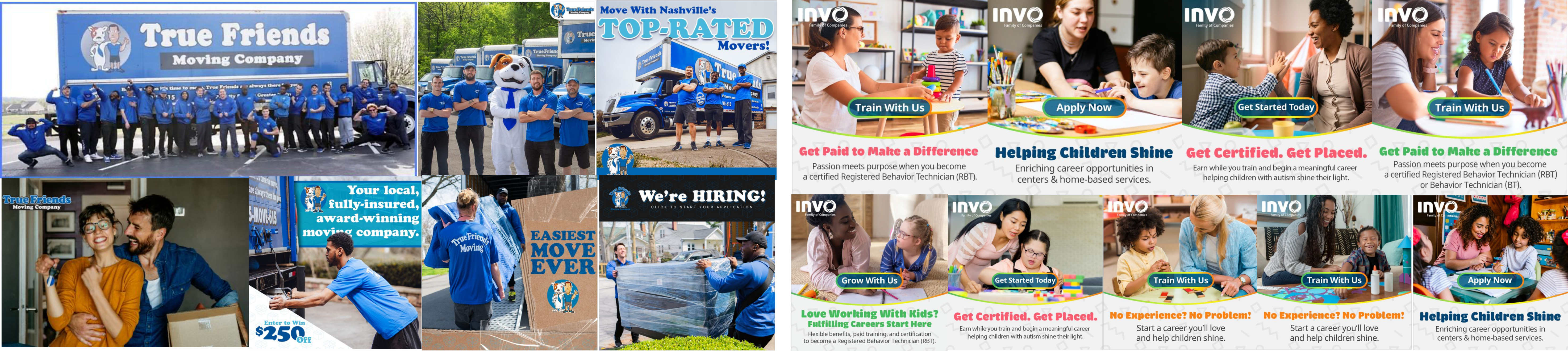}
    \caption{\texttt{True Friends Moving Company} (left): 7 distinct images containing people, all exclusively depict men as movers. \\ \texttt{Invo Healthcare} (right): 9 distinct images containing people, all exclusively depict women as healthcare staff.}
    \label{fig:anecdotal}
\end{figure*}

A search of the Ad Library easily uncovers anecdotal examples of job advertisers across many occupations using people of only one gender in the images chosen for their ads.
See Figure~\ref{fig:anecdotal} for an illustration. 
Furthermore, we found examples of advertisers who invest into creating hundreds of distinct images for their campaigns, yet an overwhelming majority of those images depict people of only one gender (see Figure \ref{fig:best_auto_service_repair}).
Motivated by these examples, we perform a comprehensive study of gendered image selection by job advertisers seeking to employ truck drivers and nurses.

\subsection{Perceived Gender Representation Among Truck Driver and Nurse Advertisers}
\label{sec:gender_discrimination}

\subsubsection{Data Collection:}
To quantify how prevalent is the selective use of gendered people in job ad images, in January 2023 we scraped data from the Facebook Ad Library and saved the job ad images for two occupations - truck driver and nurses. 
We restricted our search to the employment ads category and to only those ads which contain images and memes.
We chose these occupations because (1) they currently have a real world defacto skew towards a specific gender (91.9\% of truck drivers in the US are men, 87.9\% of nurses in the US are women \cite{bls2020labor}) and (2) we found thousands of ads for these occupations in the Ad Library, allowing for a large scale analysis. 

Since the Ad Library does not load all search results when there are more than a few hundred, obtaining data on all advertisers running employment ads for truck drivers or nurses is not merely a task of performing one search. 
Instead, to stay within the number of results that load, we 
turned each keyword search into 26 separate searches by expanding a given keyword with an additional letter of the alphabet, i.e. we
searched and downloaded results for the keywords \texttt{truck driver a} through \texttt{truck driver z} and \texttt{nurses a} through \texttt{nurses z}\footnote{A sample URL for the \texttt{truck driver a} keyword: \url{https://www.facebook.com/ads/library/?active\_status=all&ad\_type=employment\_ads&country=US&q=truck\%20driver\%20a&search\_type=keyword\_unordered&media\_type=image_and_meme}}. 
For each job advertiser whose ads appeared in the search results, we recorded their Page ID, and then performed a separate search\footnote{A sample URL for \texttt{Truck Driver Recruiting America}: \url{https://www.facebook.com/ads/library/?active_status=all&ad_type=employment_ads&country=US&view_all_page_id=106206904474383&search_type=page&media_type=image_and_meme}} 
to download all the images, excluding the videos, of all the advertiser's active ad campaigns at the time. 
We explored different variations of the keywords such as singular or plural forms and whether or not to include quotes\footnote{A search for the keyword enclosed in double quotes returns results with an exact match in the content of the ad.}, and ultimately settled on the form that produced the highest number of search results\footnote{See Appendix~\ref{sec:app:queries-urls}
for precise queries and select advertiser URLs.}.
Although our data collection may have excluded some advertisers (e.g., those whose ad text contains only the words \texttt{nurses} or \texttt{nurses} followed by digits or special characters), we believe our sample is large enough to draw conclusions. 

\subsubsection{Data Annotation:}
\begin{table}[htb!]
\caption{Data Summary of Image, Distinct Image, and People-Image Use by Truck Driver and Nurses Advertisers}
\label{tab:truck-nurse-summary}
\resizebox{\columnwidth}{!}{
\begin{tabular}{@{}l|cc@{}}
\toprule
\multicolumn{1}{c|}{\textbf{Statistic}}                                                                             & \textbf{\begin{tabular}[c]{@{}c@{}}Truck Driver \\ Advertisers\end{tabular}} & \textbf{\begin{tabular}[c]{@{}c@{}}Nurses \\ Advertisers\end{tabular}} \\ \midrule
\# of images                                                                                                        & 4,196                                                                        & 2,148                                                                  \\
\# of distinct images                                                                                               & 2,741                                                                        & 1,581                                                                  \\
\# of advertisers                                                                                                   & 446                                                                          & 360                                                                    \\
\# of excluded advertisers                                                                                          & 47                                                                           & 58                                                                     \\
\begin{tabular}[c]{@{}l@{}}\# of advertisers whose campaigns \\ contain at least one image with people\end{tabular} & 159                                                                          & 259                                                                    \\ \bottomrule
\end{tabular}
}
\end{table}
Since an advertiser may re-use the same image in many campaigns, for each advertiser, we considered both the total number of images used and the number of distinct images used. We determined whether two ad campaign images are distinct using a hashing algorithm of the open source \texttt{imagededup} tool~\cite{idealods2019imagededup}.

We annotated each image for whether it contains people and if so, the perceived gender of the people in it. We did not consider people in the images who were not clearly representatives of the occupation being recruited for (e.g., children who the nurse was attending to, or family of a truck driver waiting for them at home). Further, we corroborated our results by recruiting crowdworkers through Amazon Mechanical Turk (MTurk) who were paid above the federal minimum wage. As a result, each distinct image was annotated by 2 distinct annotators. We excluded from analyses the advertisers for whom we could not reach an agreement on annotation, or were unable to decisively determine the gender of all people due to occlusion, blurriness or use of avatars.  
We summarize our annotated data in Table \ref{tab:truck-nurse-summary}.

\subsubsection{Findings:} 
\begin{figure*}[htb]
    \centering
    \includegraphics[width = \textwidth]{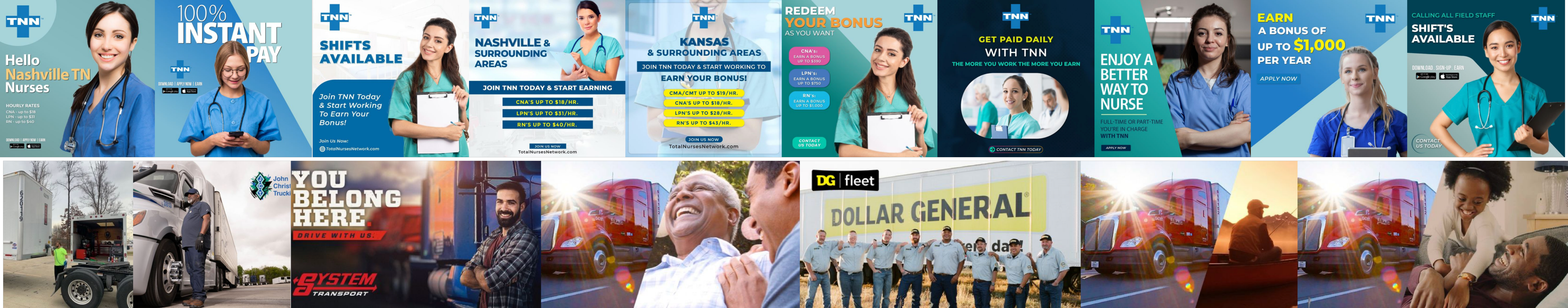}
    \caption{Example advertisers with at least $10$ distinct images depicting only people of stereotypical gender.  \texttt{Total Nurses Network} (top row) depicts women in all 10 images. \texttt{HireMaster} (bottom row) depicts men in all 7 images.}
    \label{fig:hiremaster_totalnurse}
\end{figure*}

\begin{table*}[htb!]
\caption{A Breakdown of Advertiser Practices Based on the Number of Distinct Images Used}
\label{tab:truck-nurse-distinct}
\resizebox{0.85\textwidth}{!}{
\begin{tabular}{c|ccc|ccc}
\hline
\multirow{3}{*}{\textbf{\begin{tabular}[c]{@{}c@{}}\# of Distinct \\ Images\end{tabular}}} & \multicolumn{3}{c|}{\textbf{\% (\#) of Truck Driver Advertisers}} & \multicolumn{3}{c}{\textbf{\% (\#) of Nurses Advertisers}} \\ \cline{2-7} 
 &
  \textbf{Total} &
  \textbf{\begin{tabular}[c]{@{}c@{}} using $\geq 1$ image \\ of a person \end{tabular}} &
  \textbf{\begin{tabular}[c]{@{}c@{}} whose images contain \\ only men\end{tabular}} &
  \textbf{Total} &
  \textbf{\begin{tabular}[c]{@{}c@{}} using $\geq 1$ image \\ of a person  \end{tabular}} &
   \textbf{\begin{tabular}[c]{@{}c@{}} whose images contain \\ only women\end{tabular}} \\ \hline
1                             & 187        & 18\% (33)        & 76\% (25)        & 92      & 70\% (64)      & 70\% (45)   \\
2-5                           & 133        & 44\% (59)        & 71\% (42)      & 118     & 91\% (107)     & 45\% (48)   \\
6-9                           & 35         & 80\% (28)        & 46\% (13)    & 53      & 92\% (49)      & 16\% (8)     \\
10-14             & 13         & 77\% (10)        & 60\% (6)       & 20      & 100\% (20)      & 30\% (6)     \\ 
$\geq 15$ & 31 & 94\% (29) & 14\% (4) & 19 & 100\% (19) & 5\% (1) \\ \hline
\textbf{Total} & 399 & 40\% (159) & 57\% (90) & 302 & 86\% (259) & 42\% (108) \\ \hline
\end{tabular}}
\end{table*}
We find that discrimination through image selection is a prevalent practice among the advertisers we study. 
Specifically, among 159 truck driver advertisers whose campaigns contain people, 90 or 57\% depict only men as truck drivers.
Among 259 nurses advertisers whose campaigns contain people, 108 or 42\% depict only women as nurses.
Although we observe that the fraction of advertisers using images depicting only people of the stereotypical gender for their profession typically decreases as the number of distinct images used goes up; we find examples of extensive stereotypical image selection across all ranges of the number of images per advertiser. 
Specifically, among the 33 truck driver advertisers who use only 1 image, 25 or 76\% depict a man; whereas among the 28 advertisers who use 6-9 distinct images, 13 or 46\% depict only men. For the 29 advertisers in our dataset each using at least 15 distinct images, only 4 (but still four and not zero), or 14\% depict only men. 
See Table~\ref{tab:truck-nurse-distinct} for the detailed statistics broken down by the number of distinct images used and Figure~\ref{fig:hiremaster_totalnurse} for examples of stereotypical image selection by two advertisers. 

\paragraph{Detailed Analysis of Advertisers Who Put Deliberate Effort in Their Image Selection} 

We now restrict our analysis to the advertisers who use 15 or more distinct images across their campaigns, with at least 5 of those containing people, thus focusing on 22 truck driver and 18 nurses advertisers.
We use the number of distinct images as a proxy for deliberate effort and thought the advertiser put into the visual messaging of their ad campaign, including thought towards the questions of diversity. 

For each such advertiser, we present the raw statistics on the images used, and compute the fraction of images that depict men (for truck driver advertisers) or women (for nurses advertisers) among all distinct images depicting people. 
See Tables~\ref{tab:truck_15_distinct_5_men} and~\ref{tab:nurse_15_distinct_5_women} in Appendix~\ref{sec:app:gender-race-data}.
Analyzing the image use by these presumably more deliberate advertisers, we find only two who exclusively use images of one gender.
However, the images are still dominated by the stereotypical gender choices -- 415/503 (83\%) of images depicting people used by truck driver advertisers use men and 270/380 (71\%) of images depicting people used by nurses advertisers use women.

\begin{figure}[hb!]
    \includegraphics[width=0.8\columnwidth]{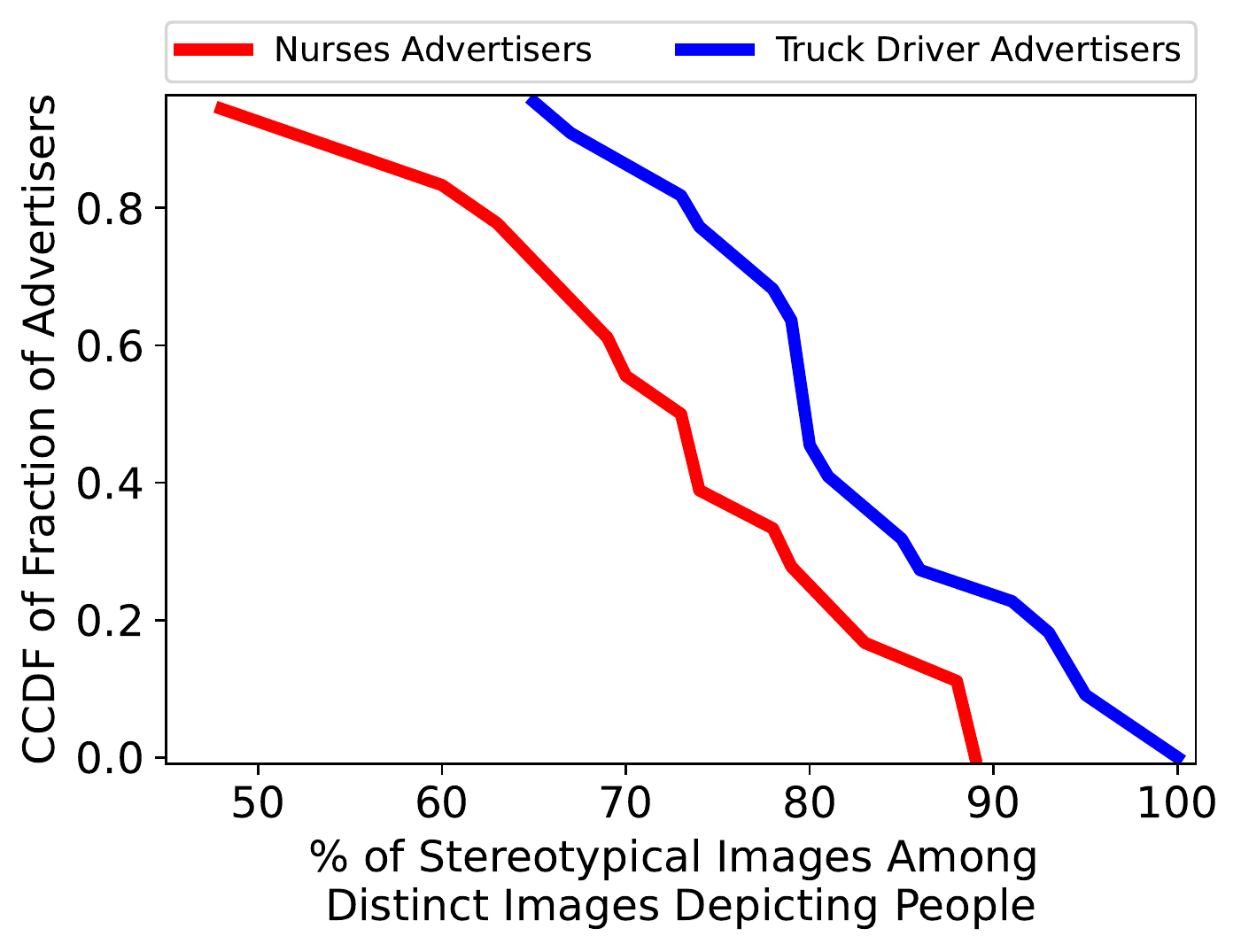}
    \caption{CCDF of the fraction of advertisers for a given percentage of stereotypical images use.}
    \label{fig:reverse_cdf}
\end{figure}

In Figure~\ref{fig:reverse_cdf} we plot the fraction of advertisers whose percentage of stereotypical image use among distinct images depicting people exceeds a certain value; i.e. the complementary cumulative distribution function (CCDF) of advertisers as a function of the percentage of stereotypical image use.
As can be seen from it, the stereotypical image use is prevalent among these advertisers.
For example, for 15/22 (68\%) truck driver advertisers and 6/18 (33\%) nurse advertisers, the percentage of stereotypical image use exceeds 78\%. 
Overall, the average rate of stereotypical gender use across all advertisers is $0.73$ for truck drivers and $0.82$ for nurses. 

As an aside, we observe that truck driver and nurses advertisers differ in their frequency of depicting people in their images. Truck driver advertisers feature more distinct images, but many of them depict trucks without people (the average fraction of images containing people across all truck driver advertisers is 0.35); whereas nurses advertisers may use fewer distinct images, but a larger fraction of them contain people (average fraction -- 0.77).

In summary, we find that even among the advertisers who have invested effort into the selection of images for their ads, the selective image use resembling discriminatory targeting is prevalent.

\paragraph{Diverse Outreach}
We now search for examples of advertisers who may be deliberately trying to diversify their workforce by exclusively depicting women in truck driver and men in nurse ads. 
We find limited evidence of such efforts -- 5/159 truck driver and 4/259 nurses advertisers. 
Furthermore, most of these advertisers use a single distinct image, which limits our ability to judge whether the non-stereotypical image selection was deliberate. 
We highlight \texttt{DHS Logistics Solution} in Figure \ref{fig:dhs_logistics_solution} of Appendix \ref{sec:app:gender-race-data} as an exception. 

\subsection{Perceived Gender and Race Representation Among Select Advertisers}
With the goal of studying whether advertisers may be selecting images to discriminate by race in addition to gender, we now focus our analysis on a select set of advertisers and study the depiction of people in their images according to both gender and race. 

\paragraph{Data Collection:}
We chose a diverse set (in terms of worker hours, domains, skill sets, demographic distribution, and income levels) of occupations and associated advertisers to study based on  Bureau of Labor Statistics (BLS) data from 2021 \cite{bls2020labor}, contingent worker supplement from 2018 ~\cite{bls2018contingent} and availability of job ads in the Ad Library containing images of people, leading to the following:\\
\noindent \textit{Occupation Specific Advertisers (N=15): } \texttt{BestBuy, Doordash, Eataly, Geico Careers, Drive with HopSkipDrive, Instacart, Drive with Lyft, Nationwide Job Search for Education, Nationwide Job Search for Information Technology, Nurse Recruiter, NYPD Recruit, Safeway, TSA, Uber, UPS Jobs.} \\
\noindent \textit{Job Aggregator Advertisers (N=3):} \texttt{Monster, SimplyJobs, Talent.}
We map occupation specific advertisers to the closest BLS categories based on the type of ads; e.g., \texttt{NYPD Recruit} is mapped to the BLS category ``Police officers''.

We scraped ads of the chosen advertisers from the Ad Library in October - November 2021 (and repeated the scrape in October 2022).
See Table~\ref{tab:company-stats} and Table~\ref{tab:gender-race-pageid} in Appendix~\ref{sec:app:gender-race-data} for a detailed summary of per-advertiser statistics in 2021.

\paragraph{Data Annotation:}
We recruited U.S. based Amazon Mechanical Turk (MTurk) workers, and compensated them above the federal minimum wage to annotate all images according to perceived gender and race. 
For each image, 2 workers were asked to count, if they exist, the number of people in the image according to gender (man, woman) and race (White, Black, Asian, Hispanic). We included form logic to ensure annotations were meaningful before they were submitted. 
We included an ``Other'' option for the annotators if it wasn't possible to determine gender or race or both ($\sim$9\% of total annotations). 
For this annotation task, we asked annotators to count all people in the images, and not just employees of the occupation being advertised.
The obtained the final count of people in the image by averaging the annotations of the two annotators.  
As in the previous analysis, our method makes the simplifying assumption of binary gender for the sake of comparison with the binary BLS data. 
Furthermore, it does not annotate the images by skin tone \cite{buolamwini2018gender} since such annotation would make a comparison with BLS data impossible. 
We restrict our analyses to the two most prevalent races in the data: Black and White. 

\paragraph{Analysis:} For a specific occupation, we judge whether an advertiser may be trying to skew their ads, to discourage or attract people of a specific demographic, by comparing the rate of representation
of gender (woman) and race (Black, White) of people depicted in the images the advertiser uses with that of the U.S. workforce (given by BLS) for that occupation. If the rates of representation are different, we say that there is a \textit{deviation in representation} and indicate whether it is an over- or under-representation. 

For a given advertiser and across all their images, we compute the rate of representation of people as follows:

\aptLtoX[graphics=no,type=html]{{\small\begin{align*}    
    \text{\% Women} &= \frac{\text{# of Women}}{\text{# of Women} + \text{# of Men} + \text{# of Other}} \times 100 \\
    \text{\% White (Black)} &= \frac{\text{# of White (Black)}}{{\text{# of White} + \text{# of Black} + \text{# of Asian}} {+ \text{# of Hispanic} + \text{# of Other}}} \times 100   
\end{align*}
}}{{\small\begin{align*}    
    \text{\% Women} &= \frac{\text{\# of Women}}{\text{\# of Women} + \text{\# of Men} + \text{\# of Other}} \times 100 \\
    \text{\% White (Black)} &= \frac{\text{\# of White (Black)}}{\splitfrac{\text{\# of White} + \text{\# of Black} + \text{\# of Asian}} {+ \text{\# of Hispanic} + \text{\# of Other}}} \times 100   
\end{align*}}}

\subsubsection{Findings}
\label{sec:gender_race_findings}
\begin{figure}[htb]
    \includegraphics[width=0.95\columnwidth]{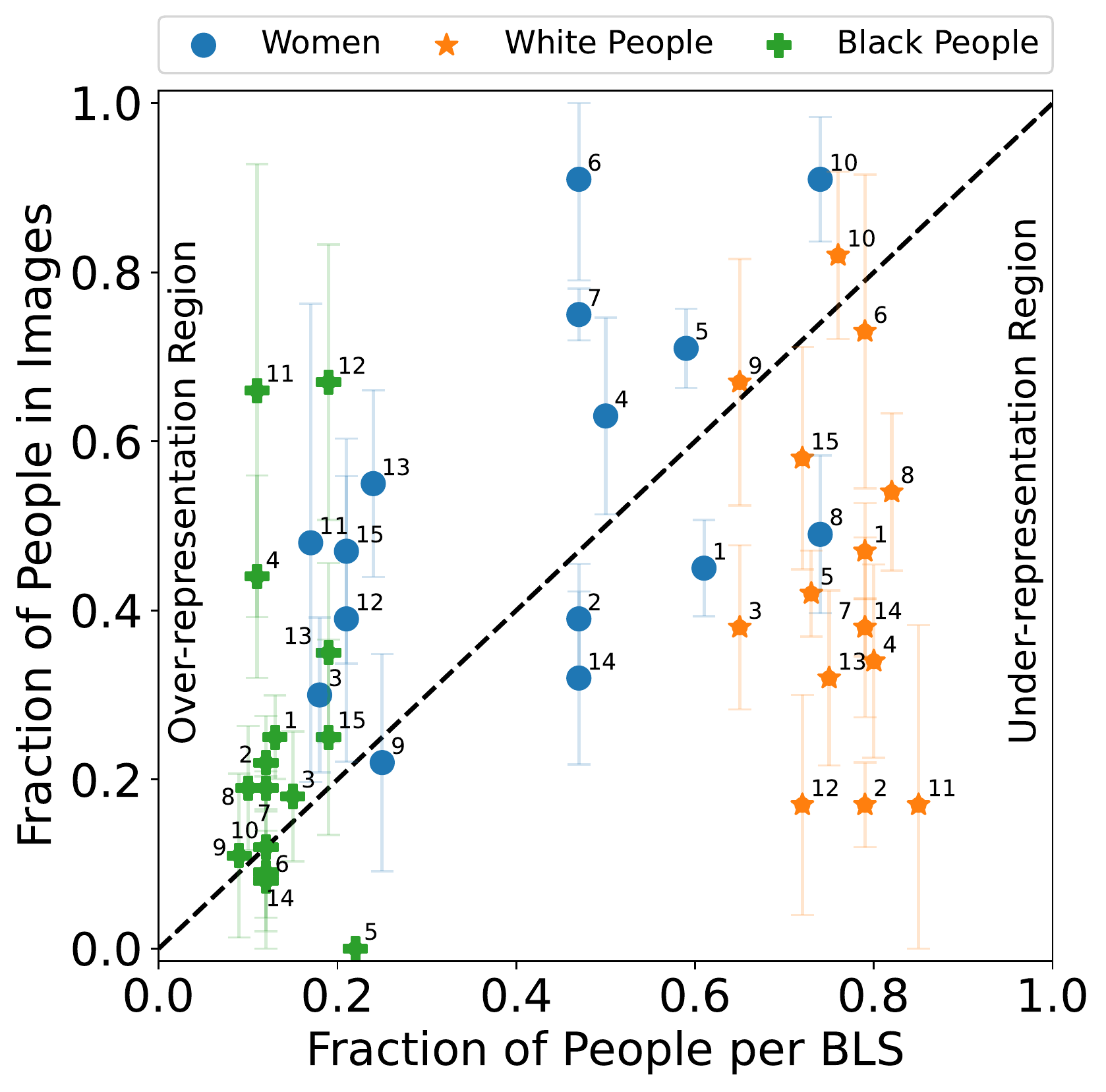}
    \caption{Rate of representation of gender and race in job ad images of select advertisers (N=15) compared to the rates of their corresponding BLS occupations. Column 1 of Table \ref{tab:company-stats} in Appendix~\ref{sec:app:gender-race-data} provides the mappings between numbers and advertiser names.}
    \label{fig:bls_comparison}
\end{figure}

Figure \ref{fig:bls_comparison} compares the rate of representation by gender (women) and race (White and Black) of people used in job ad images to the rate for that demographic characteristic per corresponding BLS occupation for each advertiser. 
Specifically, for a particular demographic characteristic and advertiser, the X-axis represents the fraction of people with that characteristic for that advertiser's occupation per BLS, and the Y-axis represents the fraction of people in the job ad images.
We plot a 45-degree reference line to visually assess the magnitude of over- or under-representation given by the regions above and below the reference line, respectively. 
Full advertiser specific statistics are presented in Table~\ref{tab:company-stats} in Appendix \ref{sec:app:gender-race-data}.

Across the 15 advertisers in Figure \ref{fig:bls_comparison} and columns 6-8 of Table~\ref{tab:company-stats}, we observe that most advertisers over-represent women and Black people, and under-represent White people. We draw definitive conclusions for advertisers for whom the error bars do not cross the reference line, finding the following:\\
\textit{Women:} 10 (over), 4 (under), 1 (inconclusive);\\
\textit{White People:} 0 (over), 12 (under), 3 (inconclusive);\\
\textit{Black People:} 8 (over), 1 (under), 6 (inconclusive). 

Our findings indicate that advertiser selection of people in job ad images with regards to gender and race is  complex and varies across different occupations.
Broadly, across a diverse set of occupations we find that some advertisers (e.g., \texttt{NYPD Recruit} and \texttt{TSA}) may be attempting to diversify their existing workforce by including more women and Black people than are employed in those professions per BLS in their images.

\paragraph{Evidence for Proactive Advertiser Selection by \texttt{Monster}:}
\begin{figure*}[hbt!]
\centering
\begin{subfigure}{.4\textwidth}
   \centering
    \includegraphics[width=0.9\linewidth]{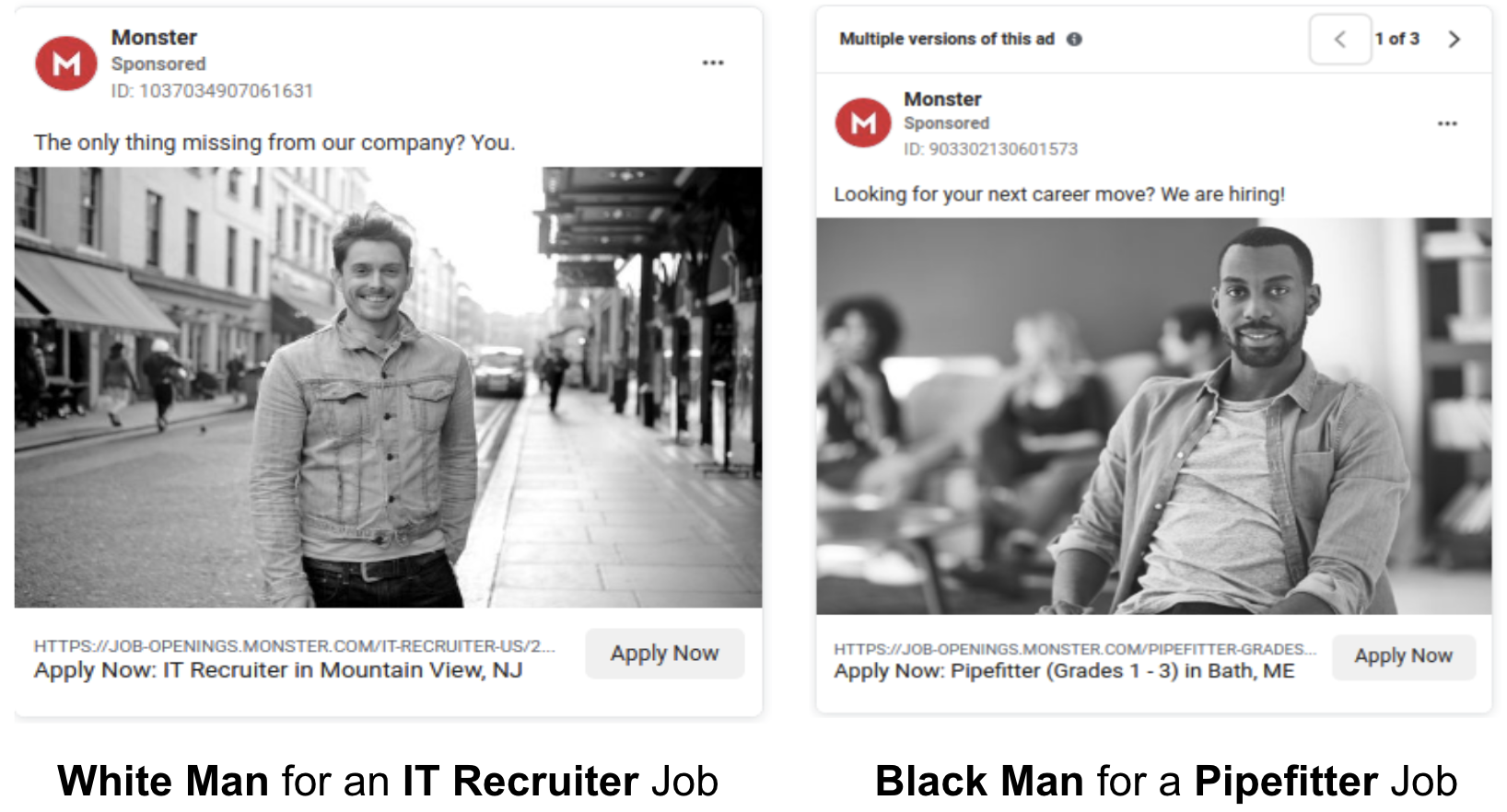}
    \caption{Stereotypical Image Use in 2021}
    \label{fig:monster_2021}
\end{subfigure}%
\begin{subfigure}{.4\textwidth}
  \centering
    \includegraphics[width=0.9\linewidth]{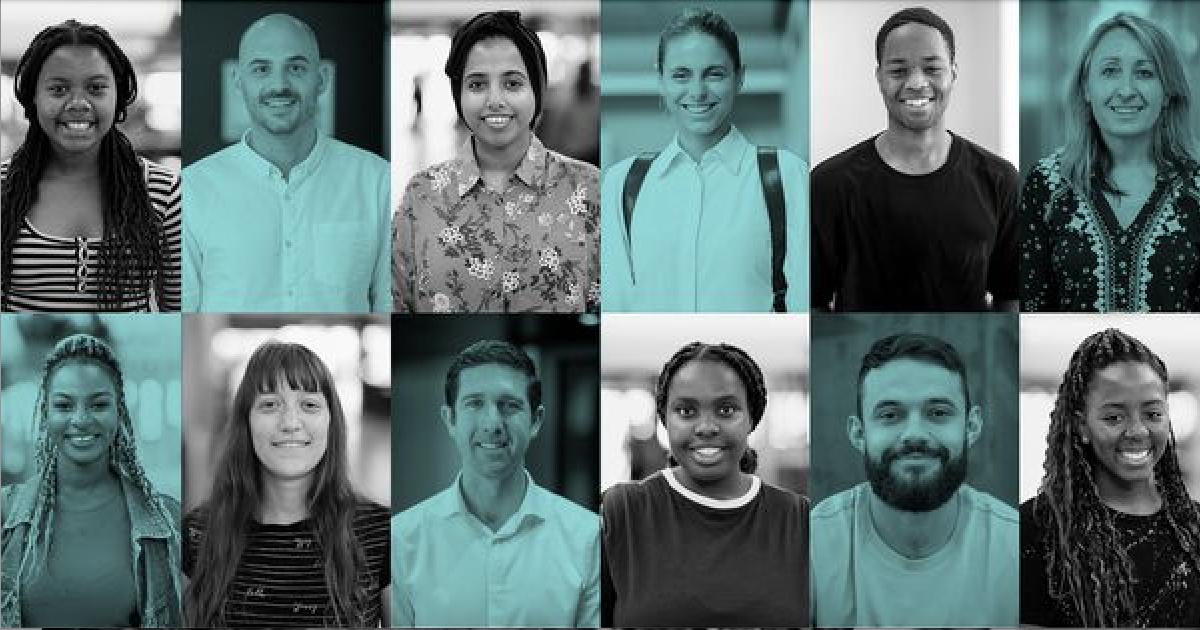}
    \caption{Diverse Image Use in 2022}
    \label{fig:monster_2022}
\end{subfigure}
\caption{Image selection by \texttt{Monster.com} across 2021 and 2022}
\end{figure*}

Advertisers may intentionally vary image selection based on their specific needs over time. To test this hypothesis, we studied  \texttt{Monster.com} (a job aggregator), which had the most number of campaigns, and compared image use in 2021 and 2022. 
We find evidence of deliberate image use and variation over time across different occupations.

\begin{table*}[htb]
\footnotesize
\caption{Percentage of Women, White and Black People with 95\% confidence interval for top three occupations of \texttt{Monster.com}}
\label{tab:monster-gender-race}
\begin{tabular}{@{}l|cc|cc|cc@{}}
\toprule
\textbf{Occupation} & \multicolumn{1}{l}{\textbf{\% Women in 2021}} & \multicolumn{1}{l|}{\textbf{\% Women in 2022}} & \multicolumn{1}{l}{\textbf{\% White in 2021}} & \multicolumn{1}{l|}{\textbf{\% White in 2022}} & \multicolumn{1}{l}{\textbf{\% Black in 2021}} & \multicolumn{1}{l}{\textbf{\% Black in 2022}} \\ \midrule
engineer            & 45 $\pm$ 7                                        & 66 $\pm$ 7                                        & 28 $\pm$ 6                                        & 44 $\pm$ 7                                        & 48 $\pm$ 7                                        & 42  $\pm$ 7                                      \\
security officer    & 49 $\pm$ 5                                        & 66 $\pm$ 3                                         & 33 $\pm$ 5                                        & 45 $\pm$ 3                                        & 50 $\pm$ 5                                        & 42 $\pm$ 3                                       \\
technician          & 28 $\pm$ 9                                       & 67 $\pm$ 13                                         & 64 $\pm$ 10                                       & 44 $\pm$ 13                                        & 21 $\pm$ 8                                       & 43 $\pm$ 13                                        \\ \bottomrule
\end{tabular}
\end{table*}

In \textit{2021}, we find evidence of stereotypical image use across occupations. 
To enable this analysis, we extracted the occupation from the destination page URL of the ads, and used it to group our findings according to the top 3 roles - security officer, engineer, and technician. Our results are presented in Table \ref{tab:monster-gender-race}. We find that the representation of women and Black people was mostly balanced for the engineer (45\% women, 48\% Black) and security officer (49\% women and 50\% Black) roles, but was skewed for technicians (28\% women and 21\% Black). We observed a White majority only in the case of technicians (64\%). 

Next, in order to delve deeper into possible advertiser intent behind image selection, we performed a reverse Google Image Search of the images. We observed that the images were generic and not present on the corresponding employer pages, implying a possible deliberate attempt by \texttt{Monster.com} to discriminate through image selection. Examples of such stereotypical images are in Figure \ref{fig:monster_2021}. An expanded list of such occupations with stereotypical image use include: Black Man - pipe fitter, White Man - IT recruiter, Black Woman - security officer, White Woman - Sr. Software Engineer, Lactation Consultant. 

In \textit{2022}, in contrast to the stereotypical image selection in 2021, most ads contained a single image with multiple people of diverse demographic characteristics as in Figure \ref{fig:monster_2022}. As a result, the percentage of women, White and Black people remains consistent across occupations as is evident in Columns 3, 5, 7 of Table \ref{tab:monster-gender-race}.

\subsection{Limitations and Ethical Considerations}
\label{sec:limitations}
The limitations of our study are a result of the data and features of the Ad Library and our analyses' choices. 
The most significant limitation due to Ad Library's capabilities is the inability to quantify the real-world impact of the selective image use, as it does not provide information about a job ad's budget, optimization objective, or the number of people reached or engaged by the ad. 
We discuss desiderata for greater platform transparency in the context of discrimination through image selection by job advertisers in Section~\ref{sec:transparency}. 
The limitations due to the choices we made are:

\textit{Selection Bias:} 
Our analyses are limited to the advertisers we chose to study and the timing of data collection; choices that may be particularly impactful for our study of racial bias in Section \ref{sec:gender_race_findings}. The selections are driven by lack of functionality that would allow for a comprehensive or representative sampling by occupation.

\textit{Annotation Bias:} Since we specified filtering criteria for MTurk workers who annotated the scraped images, the cohort of annotators may have been biased. Further, although we included an ``Other'' category when workers were not able to accurately annotate an image, they could have misreported or failed to report their perceptions. Finally, since crowdworkers operate within a social network of other crowdworkers, the mistakes could have been exacerbated by mutual influence and collaboration \cite{gray2016crowd}.

\textit{Limited View of Gender and Race:} We restrict analyses to binary gender to allow for comparison with publicly available U.S. workforce data. Further, we restrict our analysis to the two most prevalent races in the scraped data. 
We view this limited categorization as only a first step towards more nuanced analyses. 

\textit{Base Rate Comparison:} 
Several possible base rates of demographic distributions by profession could be relevant for comparing the rate of representation of people, including geography-driven and employer-driven specifics. The choice of base rate influences the analysis performed and the resulting conclusions drawn from it. We chose to compare with one such base rate, given by the BLS national averages, in order to maintain consistency with prior audits related to visual representation~\cite{metaxa2021image, kay2015unequal} and due to infeasibility of accounting for possible regional variations in demographic distributions for most jobs and advertiser types.

\smallskip
\noindent \textbf{Ethics:}
All data used was publicly accessible through the Ad Library; however, since the Ad Library only shows active ads, we expect most of the data collected for this study to not be accessible after some time. 
The study was approved by our institution's IRB.
\subsection{Algorithmic Amplification}\label{sec:amplification}
Building on the discussion in Section~\ref{sec:delivery_discrimination} of the role of the ad delivery for potential discrimination, we now illustrate how even a seemingly small difference in the rate of representation of people in job ad images can lead to significant differences across demographic groups in the delivery of those ads.

\citet{kaplan2022measurement} demonstrated that the implied racial identity of a person depicted in an image affects the demographics of the audience that is shown the ad. 
For example, in their experiments, the delivery audience of a job ad in the lumber industry depicting a Black man was 55\% Black, 45\% White; whereas the delivery audience for the same ad depicting a White man was 72\% White and only 28\% Black, even when both ads were targeting a balanced audience of Black and White people, and were otherwise equivalent (see Section 6 in \cite{kaplan2022measurement}).
Consider a hypothetical advertiser running 100 identical (in terms of targeting, budget, etc.) ads for a job in the lumber industry, with only a very slight racial skew in the images used, e.g., suppose 51 of the ads use images depicting White men and 49 -- depicting Black men.
Assuming the ad delivery algorithm performs as measured in \cite{kaplan2022measurement} and racially balanced targeting, the delivery audience of this advertiser will be $51 \times 72\% + 49 \times 45\% = 58.77\%$ White and $41.23\%$ Black.
In other words, a 2\% difference in the racial representation in image selection would lead to a $17.5\%$ difference in representation at delivery due to algorithmic amplification.
If the advertiser ran 3 ads, 2 of them depicting White men and 1 depicting a Black man, then the delivery audience would be $63\%$ White and $37\%$ Black, a $26\%$ difference.\footnote{If the ad delivery algorithm amplifies delivery unequally across races, then even if an advertiser depicts White and Black men in an equal number of ads, the delivery audience would be skewed, e.g. for lumber job ads -- $58.5\%$ White, $41.5\%$ Black.}

Taken together, our empirical findings of current advertiser practices in biased image selection combined with the results on the use of implied identity by the ad delivery algorithm from prior work and the unique role of identity in images for subsequent job applications argue that selective use of people in images for job ads should receive similar scrutiny to that of targeting and delivery. 
\section{Addressing discrimination through image selection}\label{sec:solutions}
Having demonstrated the prevalence of selective choice of people in employment advertising, and its implications for discrimination thanks to ad delivery optimization and the unique role of images to elicit action, we now discuss approaches and challenges for addressing it.
We approach the question from multiple perspectives: 
(1) transparency desiderata to perform audits of selective image choice, 
(2) transparency desiderata for the ad delivery algorithm in the context of its use of implied identity,
(3) implications from parallels with practices in other domains,
(4) metrics on image selection that could be used (by Facebook or public interest groups) to ascertain a particular advertiser's discriminatory intent.

\subsection{Desiderata for Platform's Actions}
\subsubsection{Desiderata for Enhanced Platform Transparency for Job Ads and Ad Delivery Algorithm}\label{sec:transparency}
The questions posed by our study and the challenges encountered while performing it imply that, at a minimum, the Ad Library for employment ads should provide the same data as is already being provided for social issues, elections or politics ads. 
Specifically, it should enable obtaining a list of all employment advertisers (rather than require scraping workarounds we had to resort to in Appendix~\ref{sec:app:queries-urls}). 
Then for each advertiser, it should provide information on the targeting criteria and budget chosen by the advertiser for each of their ads, as well as delivery information -- the number of people shown the ad, broken down by their demographic characteristics.
To study changes in advertiser image selection practices over time, the Ad Library should include data for both active and past campaigns.
Specific to considerations for discrimination in employment targeting, the Ad Library should provide an easy way to download the images used by the advertisers, label (or ask advertisers to self-identify) the job industry of the ad, and provide aggregate information about the people who engaged with the ad, broken down by their demographic characteristics.
Furthermore, rather that aggregating the targeting data at an advertiser level or on a weekly basis as is done for political ads, employment ad data should be provided at an ad campaign level over the duration of that campaign.
This desiderata echoes the one called for by \cite{edelson2021universal, ali2021ad, imana2021auditing}, and is feasible given the existing implementation for political ads.

However, as is clear from the works related to discrimination due to ad delivery optimization, discussed in Sections~\ref{sec:delivery_discrimination} and~\ref{sec:amplification}, transparency of advertiser image selection practices is not sufficient to understand the implications of such choices, neither for public-interest researchers nor for the advertisers themselves.
We thus call for either disabling of the use of ad delivery optimization (and the opaque machine learning algorithms underlying the total value equation) for employment advertising or a significant increase in transparency of these algorithms when applied to job ads.
A viable path for this kind of transparency has recently been proposed by \citet{imana2023having}.
Specifically, their proposal is a platform-supported 
special access API that gives public-interest researchers an ability to query the values of the estimators used in the Total Value equation, and study how they vary depending on the perceived gender and race of the person in the image and the platform-inferred demographic of the potential ad recipient.
The approach of \cite{imana2023having} addresses the concerns raised by platforms in response to requests for greater accountability -- those of user privacy and protections of the platforms' proprietary code and business interests -- through the use of differential privacy and the specifics of the set-up of the auditor - platform API interaction.

\subsubsection{Platform's Guidance to Advertisers}\label{sec:text}
The platforms should take proactive action towards ensuring that advertisers do not deliberately discriminate through their image selection.
In fact, they already do so in a somewhat analogous domain of potentially discriminatory text selection by advertisers.

Similar to the case of selective image choice influencing who applies for jobs, prior social science work shows that selective use of language affects ad recipients' actions \cite{gaucher2011evidence, goldman2019soar} (e.g. removal of the word ``aggressive" from the ad text increases the number of women applicants).
Legal precedent establishes that the practice of printing classified ads in two separate columns: ``Help wanted: Male'' and ``Help wanted: Female'' violates anti-discrimination law~\cite{now1967ads, aclu2018impact, kerman1974sex, nytimes1973sex}.
Thanks to the precedent and a relative ease for identifying discriminatory language use compared to discriminatory image use, advertising platforms such as Facebook \cite{fb2023text} and LinkedIn \cite{linkedIn2023text} already have advertiser guidance about the language for creating inclusive ads, and may be enforcing them at ad review time. 
We suggest platforms extend the guidance to image use, e.g., by updating their policies and advertiser training materials. 

Furthermore, as platforms themselves gear up to deploy Generative AI-based tools to assist advertisers in their ad campaign and creative creations~\cite{fb2023genaiadvertising, fb2023aisand}, they should be mindful not to lead the advertisers to employ image selection in a discriminatory manner.

\subsection{Measuring Advertiser Intent in Image Selection}
\label{sec:reimagine}
We now ask the normative question -- under the assumption of full access to advertiser image selection and ad campaign budget allocation, what could be the metrics to establish whether the advertiser image selection is discriminatory or to encourage a change in their choices?
This is a non-trivial question, even when all ads are allocated an equal budget.
For example, since 91.9\% of truck drivers in the US are men, should a U.S.-based advertiser be permitted to select 91.9\% images containing men for their truck job ads?
If that advertiser is using 20 distinct images, and so can only achieve representation fractions that are a multiple of $5\%$, should depicting only one woman be permitted?
Or should the targeting always strive to achieve parity by gender and ensure that everyone is represented?
In other words, should the desiderata for visual representation mimic the employment statistics as they currently are or as we as a society may aspire them to be (e.g., according to the demographic distribution of the population or equal representation of all)?
If the latter, how does one determine the aspirational ratios for jobs that are hyper-local in nature (e.g., plumbing jobs), when the demographic distribution of the population varies by geographical location? 
Should the advertiser or the advertising platform bear responsibility for determining the reference ratios, especially when data on qualified people of a certain demographic in a particular region for a particular job may not be easily available?
Finally, to what extent should an advertiser be allowed to make trade-offs between costs of building a large set of representative visual assets for their campaigns with the potential, hard-to-quantify harms of using less representative assets?  

\subsubsection{Inspiration from Jury Selection:}
One approach to consider is through the adaptation of practices for ensuring fair and impartial jury to the advertising context. 
In the U.S. that means that a pool from which juries are selected should satisfy a fair cross section requirement~\cite{seltzer1995fair, gastwirth2015statistical} via a Duren Test~\cite{1979duren}.
It measures the extent to which the jury pool demographics differ from those of the community, where the relevant community consists of individuals who are eligible for jury service.
Several disparity metrics are used to compute the differences in demographics, and case law establishes permissible levels of disparity (see \cite{jurytoolbox} for an overview). 
The challenge of adopting it in the advertising context is that the equivalents of ``eligible for the jury service" in the ``relevant community" is difficult to establish -- both who is qualified for the job and who constitutes the relevant community for a particular job (e.g., due to commute limitations) are open questions.
In the jury selection case, this data is relatively easily available through the U.S. Census.
Prior work in the computer science literature~\cite{metaxa2021image} adopts the approach of comparing the racial and gender composition of Google Image Search results to that of the composition in the U.S. workforce across specific occupations.

Literature on ensuring compliance with the Individuals with Disabilities Education Act, and determining whether disproportionalities on the basis of race exist in the placement or discipline of students with disabilities at the school district level may also be of inspiration~\cite{farkas2020district, doe2016riskratio}.
However, it also operates in the context of full information, basing calculations on the exact number of students of particular race in a particular school district that is known -- an assumption that does not hold for the advertising context.

\subsubsection{Diverse and Affirmative Outreach:}
\label{sec:affirmative}
Additional metrics may need to be established to assess the image use of advertisers who may be trying to affirmatively reach underserved populations or attempting to diversify beyond the demographics of the existing U.S. workforce for specific occupations. 
For example, the primary goal of \texttt{Black Career Network} (Page ID 70203958046) and \texttt{Black Excellence KC} (Page ID 563610990802939) is to recruit Black professionals. Therefore they could be justified in selecting many (or only) Black people for their job ad images. 
Complementarily, in our study we found examples of advertisers attempting to diversify their workforce beyond the existing U.S. workforce of specific occupations by over-representation of women or Black people, or under-representation of White people in their job ad images. 

Accounting for such practices is difficult: 
Is the choice of advertisers to empower an under-represented group through their ad images ethical or is it legal? Can this tension be reconciled if the advertiser self-identifies their reasons for doing so? How long should such practices be permitted? 
Objections have been raised to such practices in the past, e.g., 
White males have brought lawsuits for ``reverse discrimination'' against employers who have taken affirmative action to ameliorate the impacts of past discrimination \cite{eeoc1981affac}. 
The approach for accounting for such practices remains an open question, represents an ongoing policy issue \cite{fb2023vrs} and is presently the subject of active judicial scrutiny. 

\subsubsection{Judging the Outcome}
Ultimately, since the advertiser's image choice is only one of the inputs to an extremely complex and opaque system, one can argue that the desiderata should be placed on the eventual outcome of ad delivery, rather than on the targeting, and the burden of flagging potential discrimination be shared between the advertiser, the platform, and public-interest researchers.
We believe the exploration of such approaches is an important area for future work.
\section{Conclusion}
Our work highlights new representation and transparency issues at the intersection of hiring and targeted advertising.
We motivate the unique role of images in job ads on social media platforms, and then study the selective use of people in job ad images according to gender, across a nearly comprehensive set of truck driver and nurses advertisers, and a select set of advertisers according to both gender and race.

We find that a large percentage of truck driver and nurse advertisers predominantly select images containing people of stereotypical gender for that occupation, thereby, effectively, targeting or excluding people by gender.
On the contrary, we find that across a wider set of occupations, some advertisers may be selecting images to promote diversity among their job applicants.

Through its empirical and contextual findings, our study argues that the approach for ensuring non-discrimination in targeted advertising systems should look at advertiser actions more broadly than merely through the lens of the explicit targeting criteria made available by the platform. 

Through enumerating limitations of our study and their impact on the ability to judge the real-world impact of advertiser image selection, combined with discussion of implied identity use in ad delivery optimization to further amplify demographic skew, our work contributes to the growing chorus of researchers~\cite{edelson2021universal, imana2021auditing, ali2019discrimination, ali2021ad, yee2021image} and policy makers~\cite{pata2021press} advocating for increased transparency of both the advertiser practices and the platform's ad delivery algorithms, particularly in the employment domain.

Our findings raise important questions for future work in terms of the metrics that could be used for judging advertiser intentions in image selection, strategies for non-discriminatory image selections, and their potential interactions with the ad delivery algorithm.

\begin{acks}
We thank the anonymous reviewers for their constructive feedback and Judith Swan for feedback to improve our writing. 
This work was supported in part by NSF awards \#1916153, \#1956435,  \#1943584.
\end{acks}

\balance
\bibliographystyle{ACM-Reference-Format}
\bibliography{references_bibtex}
\newpage
\appendix
\section{Appendix}\label{sec:appendix}
\subsection{Download of Truck Driver and Nurses Advertiser Data}\label{sec:app:queries-urls}

\subsubsection{Search Query Keywords for Obtaining the List of Advertisers}
\label{sec:queries}
\begin{itemize}
    \item Truck Driver:  \texttt{truck driver b, ...truck driver j,truck driver l, ... truck driver z}\\ Further, we narrowed our queries and searched through keywords containing all two letter combinations for the letters \texttt{a} and \texttt{k}, since the number of search results for the single letters were too many for the page to load completely.\\ The specific keywords include: \texttt{truck driver aa, truck driver ab, ... truck driver az} and \texttt{truck driver ka, truck driver kb, ..., truck driver kz}
    \item Nurses: \texttt{nurses a, ..., nurses z}
\end{itemize}

\subsubsection{Creation of Advertiser URL for Scraping}
\label{sec:adv_url}
For each unique advertiser in the search results, we obtained the Page ID and created an advertiser specific URL which was used to scrape the respective pages.\\
The Base URL: \texttt{https://www.facebook.com/ads/library/?}\\
URL Parameters: \begin{verbatim}    
        { 
            'view_all_page_id' : page_id,
            'search_type':'page',
            'media_type':'image_and_meme',
            'country':'US',
            'active_status':'all',
            'ad_type':'employment_ads'
        }
\end{verbatim}

Sample URLs:
\begin{itemize}
    \item \texttt{CDLLife Jobs}: \url{https://www.facebook.com/ads/library/?view_all_page_id=201384326737496&search_type=page&media_type=image_and_meme&country=US&active_status=all&ad_type=employment_ads}
    \item \texttt{Total Nurses Network}: \url{https://www.facebook.com/ads/library/?view_all_page_id=85406439698&search_type=page&media_type=image_and_meme&country=US&active_status=all&ad_type=employment_ads}
\end{itemize}

\subsection{Additional Data and Analysis of Job Advertisers Image Selection}
\label{sec:app:gender-race-data}
\subsubsection{Job Advertisers Analyzed According to Gender and Race} 

\paragraph{Data Collection Period:}
The identified ads were scraped from the Ad Library during October - November 2021. We also scraped data one year later, in October 2022 to perform a longitudinal analysis. However, compared to 2021, we observed many fewer advertisers (13/18) and each advertiser had significantly fewer ad campaigns and in-turn very few images to facilitate any meaningful analysis. This may be a result of the prevailing economic conditions in 2022 - a general fall in ad sales on Facebook and stiff competition from TikTok \cite{ads2022revenue}. Therefore, other than for \texttt{Monster.com}, we restrict our analysis in this paper to ads which were run in 2021.

\paragraph{Standard Errors:}
We calculate the 95\% confidence interval of the percentages in Table \ref{tab:company-stats} according to $ p \pm 1.96 \sqrt{\frac{p(1-p)}{n}}$, where $p$ is the fraction of women, White and Black people, and $n$ is the total number of annotations across all the images of an advertiser. Here, we annotated all images (not just distinct images) by 2 distinct annotators on MTurk.

\paragraph{Interannotator Agreement:}
We used Cohen Kappa to calculate the inter-annotator agreement averaged out across all annotators for each label. The kappa value for gender is 0.86 (2021), 0.72 (2022) and race is 0.64 (2021), 0.52 (2022). The moderate agreement is not surprising due to the subjective nature of perceived gender and race classification.

The decrease in the annotator agreement in 2022 compared to that in 2021, may in part be a result of advertisers using more avatars in their job ad images, rather than human subjects. The classification of perceived gender and race of avatars can be more challenging than that for human subjects.

Additional data and analysis is presented in Tables \ref{tab:company-stats}, \ref{tab:gender-race-pageid}.

\subsubsection{Job Advertisers Analyzed According to Gender}
\begin{figure*}[ht]
    \centering
    \includegraphics[width=\textwidth]{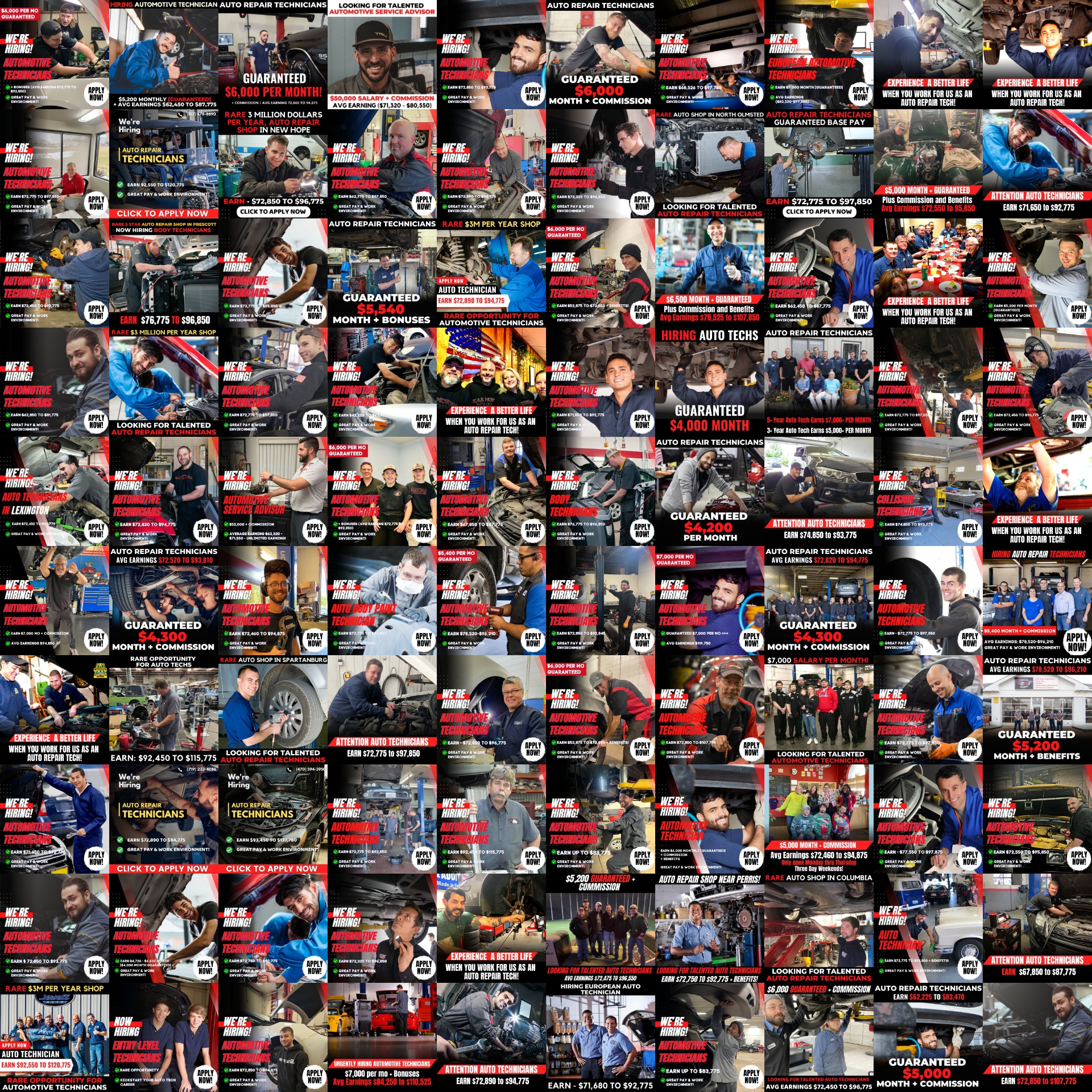}
    \caption{\texttt{Best Auto Service \& Repair} is an example of an advertiser using people of demographic characteristics similar to their existing non-diverse workforce in job ad images. Specifically, they used 100 distinct images containing people where nearly all images (93/100) exclusively depict men. This is reflective of the 2021 BLS data for ``Installation, maintenance, and repair occupations'' where women make up only 4.2\% of the current workforce.}
    \label{fig:best_auto_service_repair}
\end{figure*}

\begin{figure*}[ht]
    \centering
    \includegraphics[width=\textwidth]{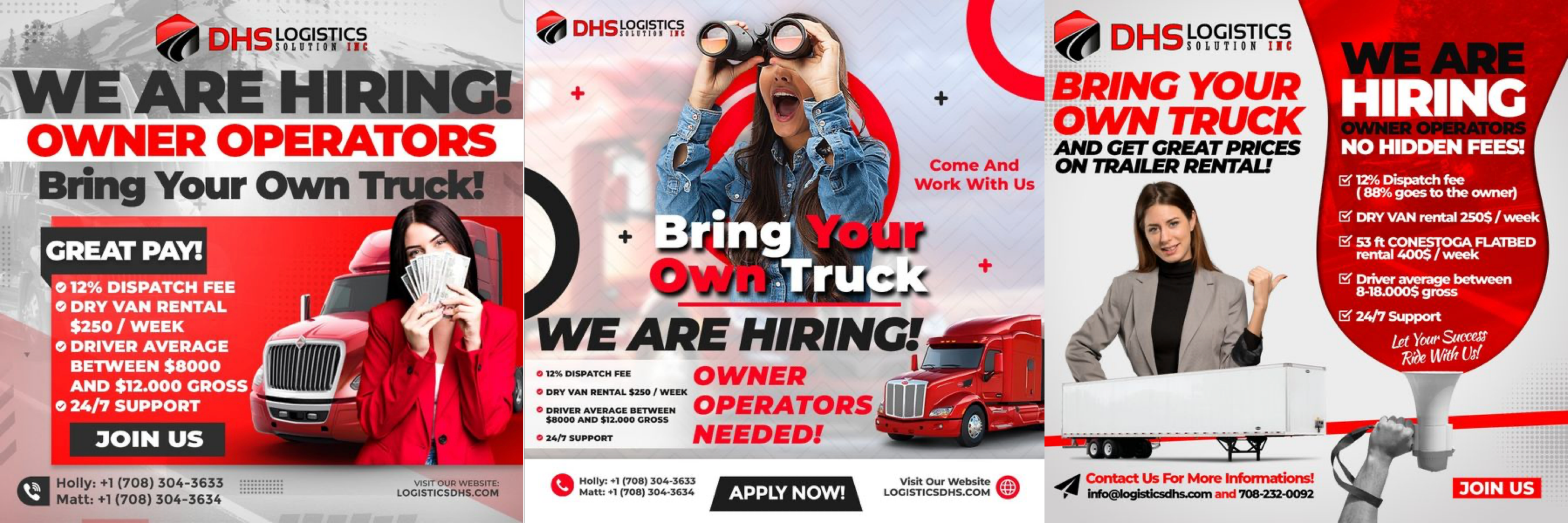}
    \caption{Diverse Outreach by \texttt{DHS Logistics Solution}: 3 out of 6 distinct images contain people and all 3 depict women.}
    \label{fig:dhs_logistics_solution}
\end{figure*}

Additional data and analysis is presented in Tables \ref{tab:truck_15_distinct_5_men}, \ref{tab:nurse_15_distinct_5_women}, and Figures 
\ref{fig:best_auto_service_repair}, \ref{fig:dhs_logistics_solution}.

\begin{table*}[tb]
\caption{ 
Columns 1-5: Occupation-specific Job Advertisers (Rows 2-16) and Job Aggregator (Rows 17-19) Data Summary from 2021. Columns 6-8: Percentage of Women, White and Black People Corresponding to the Respective BLS Data, Across All Job Ad Images for data obtained in 2021. The data from our analysis with 95\% confidence interval is present first followed by the BLS Data within parenthesis}
\resizebox{\textwidth}{!}{
\begin{tabular}{c|l|l|ccc|ccc}
\hline
\multicolumn{1}{c|}{\textbf{\begin{tabular}[c]{@{}c@{}}Serial\\ Number\end{tabular}}} & \multicolumn{1}{c|}{\textbf{Advertiser}}                                                   & \multicolumn{1}{c|}{\textbf{Occupation}}                                                          & \textbf{\# of Images} & \textbf{\begin{tabular}[c]{@{}c@{}}\# of Distinct\\ Images\end{tabular}} & \textbf{\begin{tabular}[c]{@{}c@{}}\# of Distinct Images\\ Containing People\end{tabular}} & \textbf{\begin{tabular}[c]{@{}c@{}}\% Women \\ (BLS)\end{tabular}} & \textbf{\begin{tabular}[c]{@{}c@{}}\% White \\ (BLS)\end{tabular}} & \textbf{\begin{tabular}[c]{@{}c@{}}\% Black \\ (BLS)\end{tabular}} \\ \hline
1 & \texttt{BestBuy}                                                                                    & \begin{tabular}[c]{@{}l@{}}Sales and Office \\ Occupations\end{tabular}                           & 147                   & 3                                                                        & 3                                                                                          & 45 $\pm$ 6 (61)                                                            & 47 $\pm$ 6 (79)                                                            & 25 $\pm$ 5 (13)                                                            \\ 
2 & \texttt{Doordash}                                                                                   & GIG                                                                                               & 108                   & 16                                                                       & 12                                                                                         & 39 $\pm$ 7 (47)                                                            & 17 $\pm$ 5 (79)                                                            & 22 $\pm$ 6 (12)                                                            \\ 
3 & \texttt{Eataly}                                                                                     & Chefs and Head Cooks                                   & 48                    & 14                                                                       & 13                                                                                         & 30 $\pm$ 9 (18)                                                            & 38 $\pm$ 10 (65)                                                            & 18 $\pm$ 8 (15)                                                            \\ 
4 & \texttt{Geico Careers}                                                                                     & Insurance Sales Agents                                 & 33                    & 12                                                                       & 11                                                                                         & 63 $\pm$ 12 (50)                                                            & 34 $\pm$ 11 (80)                                                            & 44 $\pm$ 12 (11)                                                            \\ 
5 & \begin{tabular}[c]{@{}l@{}}Drive with\\ HopSkipDrive\end{tabular}                          & Bus Drivers, School                                                                               & 181                   & 8                                                                        & 8                                                                                          & 71 $\pm$ 5 (59)                                                            & 42 $\pm$ 5 (73)                                                            & 0 (22)                                                             \\ 
6 & \texttt{Instacart}                                                                                  & GIG                                                                                               & 11                    & 4                                                                        & 4                                                                                          & 91 $\pm$ 12(47)                                                            & 73 $\pm$ 19 (79)                                                            & 9 $\pm$ 12 (12)                                                             \\ 
7 & \texttt{Drive with Lyft}                                                                            & GIG                                                                                               & 383                   & 35                                                                       & 23                                                                                         & 75 $\pm$ 3 (47)                                                            & 38 $\pm$ 3 (79)                                                            & 19 $\pm$ 3 (12)                                                            \\ 
8 & \begin{tabular}[c]{@{}l@{}}\texttt{Nationwide Job Search}\\ \texttt{for Education}\end{tabular}              & \begin{tabular}[c]{@{}l@{}}Education and Training \\ Occupations\end{tabular}                     & 55                    & 45                                                                       & 45                                                                                         & 49 $\pm$ 9 (74)                                                            & 54 $\pm$ 9 (82)                                                            & 19 $\pm$ 7 (10)                                                            \\ 
9 & \begin{tabular}[c]{@{}l@{}}\texttt{Nationwide Job Search for}\\ \texttt{Information Technology}\end{tabular} & \begin{tabular}[c]{@{}l@{}}Computer and Mathematical \\ Occupations\end{tabular}                  & 20                    & 13                                                                       & 12                                                                                         & 22 $\pm$ 13 (25)                                                            & 67 $\pm$ 15 (65)                                                            & 11 $\pm$ 10 (9)                                                             \\ 
10 & \texttt{Nurse Recruiter}                                                                            & Healthcare Practitioners                                                                          & 29                    & 8                                                                        & 8                                                                                          & 91 $\pm$ 7 (74)                                                            & 82 $\pm$ 10 (76)                                                            & 12 $\pm$ 8 (12)                                                            \\ 
11 & \texttt{NYPD Recruit}                                                                               & Police Officers                                                                                   & 6                     & 6                                                                        & 6                                                                                          & 48 $\pm$ 28 (17)                                                            & 17 $\pm$ 21 (85)                                                            & 66 $\pm$ 27 (11)                                                            \\ 
12 & \texttt{Safeway}                                                                                    & \begin{tabular}[c]{@{}l@{}}Laborers and Freight, Stock, \\ and Material Movers, Hand\end{tabular} & 16                    & 6                                                                        & 6                                                                                          & 39 $\pm$ 17 (21)                                                            & 17 $\pm$ 13 (72)                                                            & 67 $\pm$ 16 (19)                                                            \\ 
13 & \texttt{TSA}                                                                                        & \begin{tabular}[c]{@{}l@{}}Protective Service \\ Occupations\end{tabular}                         & 39                    & 11                                                                       & 11                                                                                         & 55 $\pm$ 11 (24)                                                            & 32 $\pm$ 10 (75)                                                            & 35 $\pm$ 11 (19)                                                            \\ 
14 & \texttt{Uber}                                                                                       & GIG                                                                                               & 40                    & 21                                                                       & 19                                                                                         & 32 $\pm$ 10 (47)                                                            & 38 $\pm$ 11 (79)                                                            & 8 $\pm$ 6(12)                                                             \\ 
15 & \texttt{UPS Jobs}                                                                                   & \begin{tabular}[c]{@{}l@{}}Laborers and Freight, Stock, \\ and Material Movers, Hand\end{tabular} & 27                    & 21                                                                       & 17                                                                                         & 47 $\pm$ 13 (21)                                                            & 58 $\pm$ 13 (72)                                                            & 25 $\pm$ 12 (19)                                                            \\ \hline
1 & \texttt{Monster}                                                                                    & Multiple                                                                                          & 499                   & 104                                                                      & 102                                                                                        & 44 $\pm$ 3                                                                 & 36 $\pm$ 3                                                                & 37 $\pm$ 3                                                                \\
2 & \texttt{SimplyJobs}                                                                                 & Multiple                                                                                          & 104                   & 89                                                                       & 45                                                                                         & 45 $\pm$ 7                                                                 & 52 $\pm$ 7                                                                 & 14  $\pm$ 5                                                                \\
3 & \texttt{Talent}                                                                                     & Multiple                                                                                          & 109                   & 77                                                                       & 38                                                                                         & 46 $\pm$ 7                                                                 & 75 $\pm$ 6                                                                & 2 $\pm$ 2                                                                 \\ \hline
\end{tabular}
}
\label{tab:company-stats}
\end{table*}

\begin{table*}[ht]
\caption{Advertiser List Analyzed According to Gender and Race Corresponding to Page ID on Facebook}
\label{tab:gender-race-pageid}
\begin{tabular}{@{}ll@{}}
\toprule
\multicolumn{1}{c}{\textbf{Advertiser}}          & \multicolumn{1}{c}{\textbf{Page ID}} \\ \midrule
\texttt{BestBuy}                                          & 12699262021                          \\
\texttt{Doordash}                                         & 534754226586678                      \\
\texttt{Eataly}                                           & 443671242427553                      \\
\texttt{Geico Careers}                                    & 52380474954                          \\
\texttt{Drive with HopSkipDrive}                          & 103480144441214                      \\
\texttt{Instacart}                                        & 369288959794283                      \\
\texttt{Drive with Lyft}                                  & 1023523827667350                     \\
\texttt{Nationwide Job Search for Education}              & 102379295254304                      \\
\texttt{Nationwide Job Search for Information Technology} & 110279211123776                      \\
\texttt{Nurse Recruiter}                                  & 112414138834087                      \\
\texttt{NYPD Recruit}                                     & 143154955728131                      \\
\texttt{Safeway}                                          & 78143372410                          \\
\texttt{TSA}                                              & 782005221949498                      \\
\texttt{Uber}                                             & 120945717945722                      \\
\texttt{UPSJobs}                                          & 93397977942                          \\
\texttt{Monster}                                          & 87877000648                          \\
\texttt{SimplyJobs}                                       & 697292896998339                      \\
\texttt{Talent}                                           & 108919760729002                      \\ \bottomrule
\end{tabular}
\end{table*}

\begin{table*}[htb]
\caption{Image selections of truck driver advertisers who use $\geq 15$ distinct images and $\geq 5$ distinct images containing people. We calculate the 95\% confidence interval of the fraction in Column 6 according to $ p \pm 1.96 \sqrt{\frac{p(1-p)}{n}}$, where $p$ is the fraction of people of stereotypical gender and $n$ is the total number of annotations across all the distinct images of an advertiser.}
\label{tab:truck_15_distinct_5_men}
\resizebox{\textwidth}{!}{
\begin{tabular}{@{}l|ccccc@{}}
\toprule
\multicolumn{1}{c|}{\textbf{Advertiser Name}} & \textbf{\# of Images} & \textbf{\begin{tabular}[c]{@{}c@{}}\# of Distinct \\ Images\end{tabular}} & \textbf{\begin{tabular}[c]{@{}c@{}}\# of Distinct Images \\ Depicting People\end{tabular}} & \textbf{\begin{tabular}[c]{@{}c@{}}\# of Distinct Images \\ Depicting Men\end{tabular}} & \textbf{\begin{tabular}[c]{@{}c@{}}Fraction of Men Images \\ Among Distinct Images\\ Depicting People\end{tabular}} \\ \midrule
\texttt{Best Driving Job}                              & 169                   & 83                                                                        & 30                                                                                         & 21                                                                                      & 0.70 ± 0.07                                                                                                  \\
\texttt{Best Truck Driver Job}                         & 315                   & 199                                                                       & 47                                                                                         & 40                                                                                      & 0.85 ± 0.03                                                                                                  \\
\texttt{Better Driver Jobs}                            & 50                    & 37                                                                        & 21                                                                                         & 17                                                                                      & 0.81 ± 0.09                                                                                                  \\
\texttt{Big Truck Driving Jobs}                        & 83                    & 57                                                                        & 20                                                                                         & 16                                                                                      & 0.80 ± 0.07                                                                                                  \\
\texttt{C R England}                                   & 72                    & 71                                                                        & 45                                                                                         & 42                                                                                      & 0.93 ± 0.04                                                                                                  \\
\texttt{CDL Job Now}                                   & 143                   & 80                                                                        & 35                                                                                         & 26                                                                                      & 0.74 ± 0.07                                                                                                  \\
\texttt{CDLLife Team Drivers}                          & 24                    & 24                                                                        & 5                                                                                          & 4                                                                                       & 0.80 ± 0.11                                                                                                  \\
\texttt{CDLLife.com}                                   & 463                   & 241                                                                       & 40                                                                                         & 31                                                                                      & 0.78 ± 0.04                                                                                                  \\
\texttt{CDLLife Jobs}                                  & 278                   & 160                                                                       & 27                                                                                         & 23                                                                                      & 0.85 ± 0.04                                                                                                  \\
\texttt{Drivers Job Choice}                            & 26                    & 25                                                                        & 14                                                                                         & 11                                                                                      & 0.79 ± 0.11                                                                                                  \\
\texttt{Findatruckerjob.com}                           & 231                   & 149                                                                       & 58                                                                                         & 53                                                                                      & 0.91 ± 0.03                                                                                                  \\
\texttt{Go Your Way Maine}                             & 19                    & 19                                                                        & 9                                                                                          & 6                                                                                       & 0.67 ± 0.15                                                                                                  \\
\texttt{Hammond Lumber Company Careers}                & 31                    & 19                                                                        & 10                                                                                         & 8                                                                                       & 0.80 ± 0.13                                                                                                  \\
\texttt{Higher Paying Driver Jobs}                     & 58                    & 53                                                                        & 18                                                                                         & 17                                                                                      & 0.94 ± 0.04                                                                                                  \\
\texttt{Hiremaster}                                    & 150                   & 87                                                                        & 8                                                                                          & 8                                                                                       & 1.00                                                                                                         \\
\texttt{HMD Trucking Inc}                              & 27                    & 26                                                                        & 20                                                                                         & 13                                                                                      & 0.65 ± 0.13                                                                                                  \\
\texttt{Knight Transportation}                         & 36                    & 34                                                                        & 32                                                                                         & 25                                                                                      & 0.78 ± 0.10                                                                                                  \\
\texttt{Nationwide Job Search For Transportation}      & 27                    & 22                                                                        & 21                                                                                         & 20                                                                                      & 0.95 ± 0.06                                                                                                  \\
\texttt{Ryder System Jobs}                             & 47                    & 33                                                                        & 26                                                                                         & 19                                                                                      & 0.73 ± 0.11                                                                                                  \\
\texttt{The Jobs Driver}                               & 56                    & 32                                                                        & 7                                                                                          & 6                                                                                       & 0.86 ± 0.09                                                                                                  \\
\texttt{Top Pay For Drivers}                           & 27                    & 19                                                                        & 5                                                                                          & 4                                                                                       & 0.80 ± 0.13                                                                                                  \\
\texttt{Ultimate Trucking Jobs}                        & 75                    & 50                                                                        & 5                                                                                          & 5                                                                                       & 1.00 \\ \hline              Total  & 2,407 & 1,520 & 503 & 415	 & Average: 0.82 \\ \hline 					
\end{tabular}}
\end{table*}

\begin{table*}[htb]
\caption{Image selections of nurses advertisers who use $\geq 15$ distinct images and $\geq 5$ distinct images containing people. We calculate the 95\% confidence interval of the fraction in Column 6 according to $ p \pm 1.96 \sqrt{\frac{p(1-p)}{n}}$, where $p$ is the fraction of people of stereotypical gender and $n$ is the total number of annotations across all the distinct images of an advertiser.}
\label{tab:nurse_15_distinct_5_women}
\resizebox{\textwidth}{!}{
\begin{tabular}{@{}l|ccccc@{}}
\toprule
\multicolumn{1}{c|}{\textbf{Advertiser Name}}  & \textbf{\# of Images} & \textbf{\begin{tabular}[c]{@{}c@{}}\# of Distinct \\ Images\end{tabular}} & \textbf{\begin{tabular}[c]{@{}c@{}}\# of Distinct Images \\ Depicting People\end{tabular}} & \textbf{\begin{tabular}[c]{@{}c@{}}\# of Distinct Images \\ Depicting Women\end{tabular}} & \textbf{\begin{tabular}[c]{@{}c@{}}Fraction of Women Images\\ Among Distinct Images\\ Depicting People\end{tabular}} \\ \midrule
\texttt{Applichat Healthcare}                           & 112                   & 68                                                                        & 48                                                                                         & 33                                                                                        & 0.69 ± 0.08                                                                                                    \\
\texttt{Ascension Careers}                              & 41                    & 34                                                                        & 31                                                                                         & 23                                                                                        & 0.74 ± 0.10                                                                                                    \\
\texttt{Camden Clark Medical Center}                    & 21                    & 20                                                                        & 18                                                                                         & 14                                                                                        & 0.78 ± 0.13                                                                                                    \\
\texttt{Children's Hospital Of The Kings Daughter's (CHKD)} & 39                    & 22                                                                        & 14                                                                                         & 11                                                                                        & 0.79 ± 0.12                                                                                                    \\
\texttt{Consumer Direct Care Network}                   & 20                    & 19                                                                        & 19                                                                                         & 17                                                                                        & 0.89 ± 0.10                                                                                                    \\
\texttt{Dartmouth Health Careers}                       & 24                    & 23                                                                        & 20                                                                                         & 12                                                                                        & 0.60 ± 0.14                                                                                                    \\
\texttt{Encompass Health Careers}                       & 62                    & 25                                                                        & 23                                                                                         & 17                                                                                        & 0.74 ± 0.12                                                                                                    \\
\texttt{Gifted Healthcare}                              & 27                    & 25                                                                        & 20                                                                                         & 14                                                                                        & 0.70 ± 0.13                                                                                                    \\
\texttt{Health Carousel Travel Nursing}                 & 25                    & 17                                                                        & 8                                                                                          & 7                                                                                         & 0.88 ± 0.11                                                                                                    \\
\texttt{John Knox Village Of Central Florida}           & 68                    & 44                                                                        & 23                                                                                         & 11                                                                                        & 0.48 ± 0.10                                                                                                    \\
\texttt{Kettering Health}                               & 50                    & 31                                                                        & 31                                                                                         & 20                                                                                        & 0.65 ± 0.12                                                                                                    \\
\texttt{Memorial Health Careers}                        & 25                    & 24                                                                        & 23                                                                                         & 19                                                                                        & 0.83 ± 0.11                                                                                                    \\
\texttt{MyMichigan Health}                              & 24                    & 20                                                                        & 19                                                                                         & 12                                                                                        & 0.63 ± 0.15                                                                                                    \\
\texttt{NuWest Travel Nursing}                          & 67                    & 37                                                                        & 24                                                                                         & 13                                                                                        & 0.54 ± 0.11                                                                                                    \\
\texttt{Providence Careers}                             & 19                    & 19                                                                        & 11                                                                                         & 8                                                                                         & 0.73 ± 0.14                                                                                                    \\
\texttt{Providence Health Services Careers}             & 24                    & 23                                                                        & 13                                                                                         & 9                                                                                         & 0.69 ± 0.13                                                                                                    \\
\texttt{The Guthrie Clinic}                             & 17                    & 16                                                                        & 16                                                                                         & 13                                                                                        & 0.81 ± 0.14                                                                                                    \\
\texttt{VNS Health}                                     & 19                    & 19                                                                        & 19                                                                                         & 17                                                                                        & 0.89 ± 0.10                                                                                                    \\ \bottomrule
Total & 684 & 486 & 380 & 270	 & Average: 0.73 \\ \hline 
\end{tabular}}
\end{table*}

\end{document}